\documentclass[namedreferences]{solarphysics}
\usepackage{graphicx}        
\usepackage{color}           
\usepackage{url}             
\usepackage{multirow}
\usepackage[optionalrh]{spr-sola-addons} 
\let\bibhang\relax
\let\citeauthoryear\relax
\usepackage{natbib}         

\begin{document}

\begin{article}

\begin{opening}

\title{Photometric and Thermal Cross-calibration of Solar EUV Instruments\\ {\it Solar Physics}}

\author{P.F.~\surname{Boerner}$^{1}$\sep
        P.~\surname{Testa}$^{2}$\sep
        H.~\surname{Warren}$^{3}$\sep
        M.A.~\surname{Weber}$^{2}$\sep  
        C.J.~\surname{Schrijver}$^{1}$
       }
\runningauthor{P.Boerner}
\runningtitle{Cross-calibration of Solar EUV Instruments}

   \institute{$^{1}$ Lockheed Martin Solar and Astrophysics Laboratory, A021S, Bldg 252, 3251 Hanover Street, Palo Alto, CA 94304 USA
                     email: \url{boerner@lmsal.com, schrijver@lmsal.com} \\
            }
   \institute{$^{2}$ Harvard Smithsonian Center for Astrophysics, 60 Garden Street, Cambridge, MA 02138, USA
                     email: \url{ptesta@cfa.harvard.edu, mweber@cfa.harvard.edu} \\
            }
   \institute{$^{3}$ Naval Research Laboratory, 4555 Overlook Ave. S.W., Washington, DC 20375, USA
                     email: \url{harry.warren@nrl.navy.mil} \\
             }

\begin{abstract}
We present an assessment of the accuracy of the calibration measurements and atomic physics models that go into calculating the SDO/AIA response as a function of wavelength and temperature. The wavelength response is tested by convolving SDO/EVE and {\it Hinode}/EIS spectral data with the AIA effective area functions and comparing the predictions with AIA observations. For most channels, the AIA intensities summed over the disk agree with the corresponding measurements derived from the current Version (V2) of the EVE data to within the estimated 25\% calibration error. This agreement indicates that the AIA effective areas are generally stable in time. The AIA 304 \AA\ channel, however, does show degradation by a factor of almost 3 from May 2010 through September 2011, when the throughput apparently reached a minimum. We also find some inconsistencies in the 335 \AA\ passband, possibly due to higher-order contamination of the EVE data. The intensities in the AIA 193 \AA\ channel agree to within the uncertainties with the corresponding measurements from EIS full CCD observations. Analysis of high-resolution X-ray spectra of the solar-like corona of Procyon, and of EVE spectra, allows us to investigate the accuracy and completeness of the CHIANTI database in the AIA shorter wavelength passbands. We find that in the 94 \AA\ channel, the spectral model significantly underestimates the plasma emission owing to a multitude of missing lines. We derive an empirical correction for the AIA temperature responses by performing differential emission measure (DEM) inversion on a broad set of EVE spectra and adjusting the AIA response functions so that the count rates predicted by the full-disk DEMs match the observations.
\end{abstract}
\end{opening}

\section{Introduction}

The {\it Atmospheric Imaging Assembly} (AIA; \citealp{lemen_atmospheric_2012}) is an array of telescopes which continuously observes the full solar disk in nine UV/EUV wavelength channels with high cadence (12 s for EUV channels and 24 s for UV) and spatial resolution ($4096 \times 4096$ pixels of 0.6 arcsec each). Its images have facilitated new understanding of numerous phenomena in solar physics, including the global structure of the magnetic field \citep{schrijver_2011_2011}, new types of waves associated with flares \citep{liu_direct_2011}, and the heating of active region loops \citep{warren_constraints_2011}.

Like earlier instruments such as SOHO/EIT \citep{dere_preflight_2000} and TRACE \citep{handy_transition_1999}, AIA uses normal-incidence multilayer mirror coatings to isolate a narrow spectral range ($\approx$ 10 \AA\ full width at half maximum) for each of its EUV channels; the central wavelengths of the channels are chosen to coincide with strong emission lines formed at different temperatures from 500,000 K to 20,000,000 K. AIA data consists of images with pixel values $p_{i}(\mathbf{x})$ where the index $i$ refers to one of the ten wavelength channels (nine UV/EUV and one visible light) and $\mathbf{x}$ refers to a location in the field of view. These pixel values are measurements of the solar spectral radiance integrated over the solid angle subtended by the pixel and the wavelength passband of the telescope channel:
\begin{equation} \label{E-wresp}
p_{i}(\mathbf{x}) = \int_{0}^{\infty} R_{i}(\lambda) \mathrm{d} \lambda \int_{\mathrm{pixel}~\mathbf{x}} I(\lambda, \theta) \mathrm{d} \theta .
\end{equation}
Here $R_{i}$ is the wavelength response function of the $i$-th channel of the telescope, with dimensions of digital number (DN) per unit flux at the aperture. It is possible to recast this measurement equation into an integral over temperature rather than wavelength by using a model of the emissivity of the solar plasma as a function of wavelength and temperature, and folding the emissivity with the wavelength response of the instrument to produce a temperature response function $K(T)$:
\begin{equation} \label{E-dem}
p_{i}(\mathbf{x}) = \int_{0}^{\infty}K_{i}(T) DEM(T, \mathbf{x}) \mathrm{d} T .
\end{equation}
Quantitative analysis of AIA data generally consists of using a set of observations to invert (or place constraints on) the spectral distribution of solar emission or the thermal distribution of plasma along the line of sight (the differential emission measure function, DEM$(T)$). In either case, accurate calibration -- that is, knowledge of the instrument response as a function of wavelength and temperature -- is essential. Relative errors in the calibration of AIA channels can result in much larger distortions in the inferred properties of the emitting region. Errors in the absolute calibration can bias the results of an analysis, and make it difficult to take advantage of observations from complementary instruments such as {\it Hinode}/EUV Imaging Spectrometer (EIS; \citealp{culhane_euv_2007}) and X-Ray Telescope (XRT; \citealp{golub_x-ray_2007}) to extend the temperature coverage and precision of the AIA observations.

The pre-flight calibration of AIA is described in \citealp{boerner_initial_2012}, along with a preliminary assessment of the accuracy of that calibration based on early on-orbit data. In this work, we describe a series of experiments to assess and improve the accuracy of the AIA wavelength and temperature response functions by cross-calibration with a number of other instruments. Section~\ref{S_wvlresp} describes the testing of the wavelength response with data from SDO/EVE and {\it Hinode}/EIS. Section~\ref{S_tresp} describes the assessment and adjustment of the emissivity function used to generate the temperature response function. In Section~\ref{S_dem} we review some of the applications of these results, including tests of differential emission measure inversion using AIA and other instruments.

\section{Wavelength Response}
	\label{S_wvlresp}
	
As noted in \citealp{boerner_initial_2012}, the wavelength response function of each channel is the product of the effective area $A_{\mathrm{eff}}(\lambda)$ (dimensions of $\mathrm{cm}^{2}$) and the gain $G(\lambda)$ (DN/photon). The effective area is the geometrical collecting area of the system, multiplied by the efficiency of each of the components (mirrors, filters, CCD, etc.) as a function of wavelength. The pre-flight calibration relied on component-level measurements of each optical element to determine the effective area and gain. The uncertainty in the wavelength response is thus the stackup of the uncertainties in the calibration of each component, which is approximately 25\%. There is additional uncertainty due to changes in the instrument response after the initial measurement due to contamination or other degradation of the instrument. These effects can be significant in the EUV, having resulted in sensitivity losses of a factor of 2 or more on some instruments.

Cross-calibration with other instruments that observe the Sun in the same wavelength channels therefore provides two important capabilites: it enables a check of the initial calibration accuracy, and it allows for tracking and correction of on-orbit changes in sensitivity. Fortunately, the AIA mission overlaps wth the operation of two EUV spectrometers suitable for cross-calibration: SDO/EVE (which measures full-Sun spectral irradiance at high cadence and moderate spectral resolution across the AIA EUV wavelength range), and {\it Hinode}/EIS (a slit spectrograph that measures the full range of the AIA 193 \AA\ channel with excellent spatial and spectral resolution).

\subsection{Comparison with SDO/EVE}
	\label{SS_aiaeve}

The EVE instrument on SDO \citep{woods_extreme_2012} makes measurements of the solar spectral irradiance from 60--1050 \AA\ with $\approx$ 1 \AA\ spectral resolution and a 10 sec cadence. While the stated absolute accuracy of EVE's calibration is 25\% \citep{hock_extreme_2012}, similar to the expected accuracy of the AIA pre-flight calibration, cross-calibration with EVE provides a number of advantages. EVE is optimized for maintaining accurate absolute calibration. It uses redundant optical elements, proxy models, and comparison with other irradiance monitors to continuously check its measurements, and annual rocket underflights to track degradation.

AIA and EVE measurements are compared as follows: the EVE spectral data (consisting of a solar spectral irradiance $E_{\mathrm{EVE}}(\lambda)$ in units of W m$^{-2}$ nm$^{-1}$) is folded through the AIA wavelength response function $R(\lambda)$ to produce a predicted band irradiance (in DN s$^{-1}$):
\begin{equation} \label{E-evepred}
B_{\mathrm{pred}} = \int_0^\infty \! E_{\mathrm{EVE}}(\lambda) R(\lambda) \, \mathrm{d} \lambda.
\end{equation}
The predicted band irradiances for each of the AIA EUV channels are computed in the EVE data processing pipeline for every observation. They are generated using the pre-flight AIA response functions \citep{boerner_initial_2012}, and included in the Level 2 EVL (extracted lines) data product. Note that the analysis presented here uses Version 2 of the EVE calibration (released in February 2011); it will be updated based on the revisions to EVE's absolute calibration included with the release of Version 3 of the EVE data in March 2013.

The predicted band irradiance is compared with the band irradiance actually observed by AIA ($B_{\mathrm{obs}}$). The observed band irradiance is found by summing all the pixels in an AIA Level 1 image (flat-fielded, dark-subtracted, and de-spiked), normalized by exposure time, and adjusted for the distance from AIA to the Sun (since the EVE L2 data is normalized to 1 AU). The ratio of the observed AIA count rate to the count rate predicted using the combination of EVE data and the AIA wavelength response function is the EVE normalization factor $F_{\mathrm{norm}}$:
\begin{equation} \label{E-fnorm}
F_{\mathrm{norm}} = \frac{B_{\mathrm{obs}}}{B_{\mathrm{pred}}}.
\end{equation}
EVE observes a larger field of view than AIA, but the amount of irradiance in the AIA bands outside of the AIA field is generally less than 1\% of the detected irradiance. Because AIA and EVE both operate continuously at a very high cadence, it is possible to compute $F_{\mathrm{norm}}$ for each AIA channel every 12 s over essentially the full SDO mission. 

\begin{figure}    
 \centerline{\includegraphics[width=0.7\textwidth,angle=90,clip=]{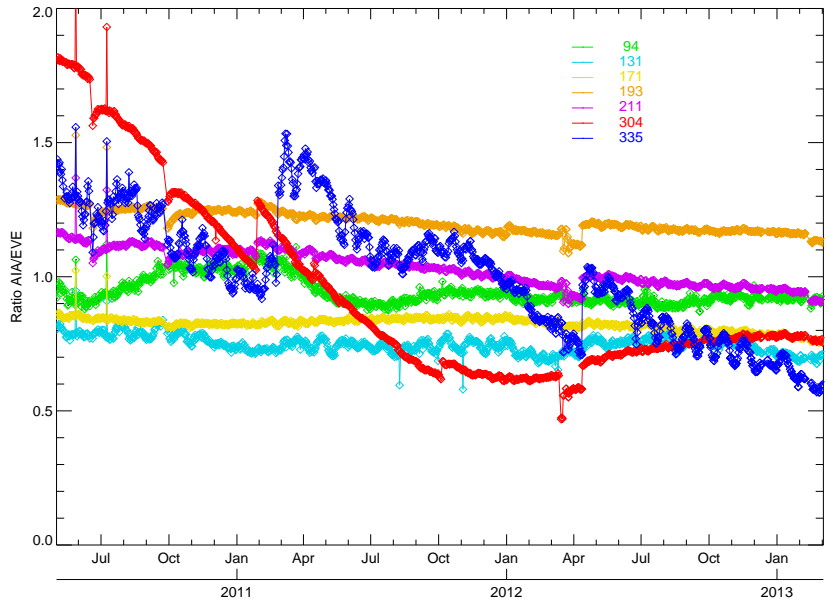}
              }
              \caption{The ratio of the total irradiance observed in each AIA EUV bandpass to that predicted by folding EVE spectra through the AIA pre-flight wavelength response functions. If we assume the EVE data are perfect, this ratio can be used as a correction factor for the AIA wavelength response. }
   \label{F_aia_eve_trend}
\end{figure}

In order to track long-term changes in the AIA sensitivity and get an overall estimate of the accuracy of the wavelength response function, it is sufficient to sample the normalization factor once per day (averaging 1 minute of AIA and EVE data). Note that EVE only operates the MEGS-B channel (used for the 370--650 \AA\ range) for a few hours per day on most days, in order to reduce the dose-dependent degradation of its sensitivity; where possible, we select the representative minute for each day from the interval when MEGS-B is operational. The results of this long-term comparison using Version 2 of the EVE calibration are shown in Figure~\ref{F_aia_eve_trend}. A number of features are immediately apparent:

\begin{enumerate}

\item For most channels, the ratio is relatively flat, or shows a slight degradation in AIA response over time (of order 5\%/year or less). The ratios on 1 May 2010, the start of normal science operations, show a DC offset from unity, indicating a discrepancy in the overall normalization of the AIA/EVE calibration. The standard deviation of the offsets in the seven EUV channels is 28\%, consistent with our estimate of the accuracy of AIA's preflight calibration.

\item There are discontinuities in the ratios whenever AIA or EVE performed CCD bakeouts (a list of the bakeouts is in Table~\ref{T_bakeouts}). EVE bakeouts generally result in a transient uncorrected increase in the EVE signal (within 1-2 weeks after the bakeout, the EVE data have been corrected for the sensitivity changes and the ratios return to their pre-bakeout trend line). AIA bakeouts produce a jump up in the ratio, which persists since the AIA data are not corrected based on these measurements. There are occasionally discontinuities when the AIA flatfields are updated ({\it e.g.} on 1 January 2012).

\item There is a long-term drop in 304 \AA\ and 335 \AA\ channel sensitivity. The 304 \AA\ degradation is particularly dramatic, although it appears to have slowed and reversed itself in September 2011. The drop is likely due to the accumulation of volatile contamination on the optics or detector telescopes. Note that the 94 \AA\ channel shares the telescope structure with the 304 \AA, and the 131 \AA\ channel with the 335 \AA; however, the typical absorption cross-section of the hydrocarbons associated with contamination is much higher at $\lambda > 300$ \AA\ than at $\lambda < 150$, so a thin layer of contamination might easily attenuate the 304 \AA\ by a factor of two without having a noticeable effect on the 94 {\AA} \citep{boerner_initial_2012}.

\item The 335 \AA\ ratio shows much greater variation on the timescale of the solar rotation (10\%) than any of the other channels (typically less than 1\%). This may indicate that the assumed shape of the 335 \AA\ wavelength response function is incorrect, causing the ratio to vary depending on the spectral distribution of the solar irradiance. However, efforts to flatten out the ratio by iteratively adjusting the wavelength response function have not enabled us to produce a realistic alternate response function that reduces the variation in the ratio while remaining compatible with the uncertainties in the instrument calibration. It is also possible that signal from higher orders in the EVE spectrum around 335 \AA\ may cause these ripples (in which case the shape of the wavelength response function may be correct).

\item The 94 \AA\ channel shows some modulation on the timescale of 1 year. This is attributable to the change in the 94 \AA\ flatfield due to burn-in by the 304 \AA\ image on their shared detector \citep{shine_flat_2010}, an effect that was not corrected for until January 2012. The CCD area corresponding to the solar disk image at 304 \AA\ has a slightly lower sensitivity at 94 \AA; thus, when SDO is at aphelion and the solar image is smallest, more of the 94 \AA\ flux (which is preferentially distributed at and above the solar limb) falls on the affected area of the detector, and thus the observed 94 \AA\ irradiance is lowest in July. 

\end{enumerate}

\begin{table}    
\caption{ History of bakeouts performed on AIA and EVE.
}
\label{T_bakeouts}
\begin{tabular}{cccc}    
  \hline                 
Date 			& Instrument	& Approximate	&	Temperature	\\
				& affected		& duration [h]	&	[$^\circ$ C]	\\
  \hline
18--Jun--2010	& EVE/MEGS		& 240 			& \\
24--Sep--2010	& EVE/MEGS		& 240			& \\
28--Jan--2011	& ATA2, 3, 4	& 2				& 10\\
25--Feb--2011	& ATA1			& 2 			& 10\\
14--Apr--2011	& ATA4			& 24			& 10\\
19--May--2011	& ATA4			& 8 			& 20\\
4--Oct--2011	& ATA4			& 36			& 20\tabnote{Heated entire telescope, not just CCD}	\\
12--Mar--2012	& EVE/MEGS		& 72			& \\
12--Apr--2012	& ATA1, 2, 3, 4	& 2				& 10\\
  \hline
\end{tabular}
\end{table}

\begin{figure}    
 \centerline{\includegraphics[width=0.7\textwidth,angle=0,clip=]{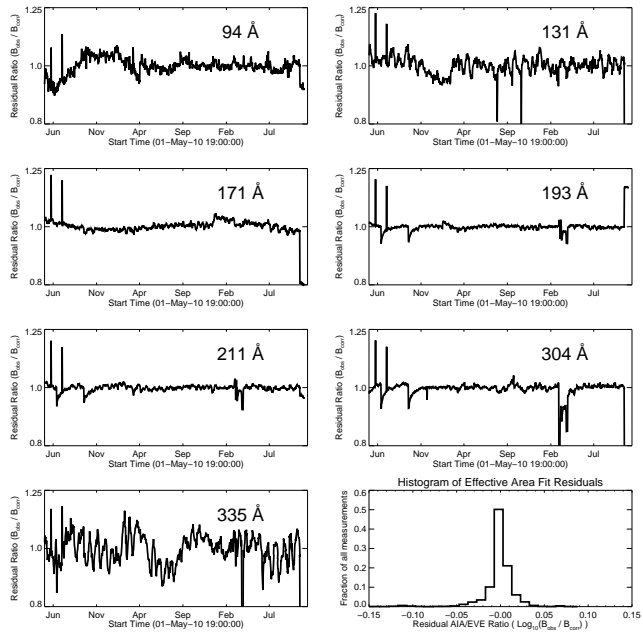}
              }
              \caption{The residual ratios left from fitting the AIA/EVE ratio in the intervals between AIA bakeouts with a flat or linear function of time. At bottom right is a histogram of the residuals from all EUV channels, showing that the vast majority of daily samples are within 2\% of the fit value.}
   \label{F_aia_eve_residuals}
\end{figure}

Some of the offset from unity and the long-term trends noted in Figure~\ref{F_aia_eve_trend} may be attributable to errors in EVE's calibration, and not in AIA's. However, since EVE is generally expected to have a better absolute calibration, and has a much better mechanism for tracking on-orbit degradation (through sounding rocket underflights), we might improve AIA's calibration by adjusting the wavelength response functions by $F_{\mathrm{norm}}$ so that the EVE-predicted band irradiances match the observations. The normalization factor is a function of time; we can approximate it as a series of polynomials for each channel and each time interval $j$ between bakeouts of that channel:
\begin{equation} \label{E-polyfit}
F'_{\mathrm{norm}}(t) = \sum_{i=0}^{n} p_{ij} (t-t_{j})^{i}.
\end{equation}
This is similar to the approach used in \citealp{hock_cross-calibration_2008} for cross-calibration of EIT and TIMED/SEE. The time-dependent approximated normalization factor is used to compute corrected predicted band irradiances $B_{\mathrm{corr}}(t) = F'_{\mathrm{norm}}(t) B_{\mathrm{pred}}(t)$. The accuracy of the correction is determined by examining the residual ratios of this fit, $B_{\mathrm{obs}} / B_{\mathrm{corr}}$ (see Figure \ref{F_aia_eve_residuals}). We find that the residual deviations from unity for all EUV channels other than 335 \AA\ are under 4\% RMS using a polynomial of order $n=0$ or 1. The polynomial coefficients $p_{ij}$ and epoch start times $t_{j}$ used to compute $F'_{\mathrm{norm}}(t)$ are included in the SolarSoftware (SSW; \citealp{freeland_data_1998} ) routine \verb+aia_get_response+, which is used to access the wavelength and temperature response functions.

\begin{figure}    
 \centerline{\includegraphics[width=0.7\textwidth,angle=0,clip=]{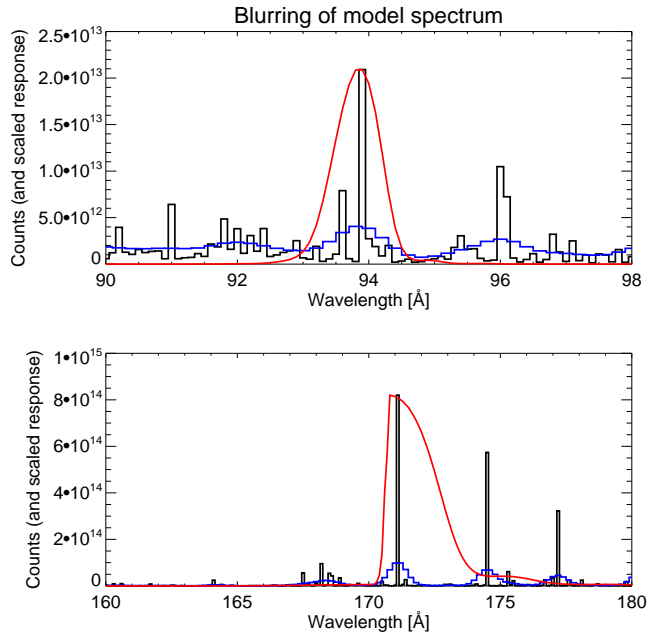}
              }
              \caption{In order to estimate the impact of EVE's spectral resolution on the comparison with AIA observations, a simulated high-resolution spectrum (black) is blurred and downsampled to match the appearance of EVE lines (blue). The blurred spectrum is folded through the AIA wavelength response (red), and the resulting count rate is compared with that predicted using the unblurred spectrum. The top and the bottom panels are for the 94 and 171 {\AA} channels, respectively.}
   \label{F_eve_resolution}
\end{figure}

The spectral resolution of EVE, while considerably higher than that of the AIA channels, may not be great enough to avoid introducing some bias into this determination of the correction factor. In order to assess this possibility, we simulated a solar spectrum at very high resolution (0.05 \AA) using the CHIANTI atomic database \citep{landi_chiantiatomic_2012}. We first folded this spectrum through the AIA wavelength response functions to produce a predicted count rate, then blurred the spectrum with a Gaussian width of 0.47 \AA\ and downsampled it to 0.16 \AA\ spectral bins (which produces a good empirical match with the appearance of the lines in the EVE Level 2 spectra around 200 \AA). We compared the count rate predicted using the blurred spectrum with that predicted by the full-resolution spectrum (Figure \ref{F_eve_resolution}). In most cases, the differences were less than 1\%; however, for the 171 \AA\ channel (where there is a strong solar emission line from Fe {\sc ix} next to the sharp Al L-edge in the response function) the slight blurring was enough to reduce the predicted count rate by approximately 10\%. For the 94 \AA\ channel (which is very narrow), the effect was an underprediction of 30--40\% (depending on the relative strength of the Fe {\sc xviii} line). This implies that, while the agreement between AIA and EVE appears to be quite good in the 94 \AA\ channel, it is possible that the assumed effective area for this channel is too high (the calibration error may be compensating for the effect of EVE's spectral resolution).

Note that a similar effect applies when attempting to fold the EVE EVS Level 2 spectral data through the response functions to reproduce the band irradiances reported in the EVL line products: the Level 2 spectra are rebinned to a slightly coarser grid than the unpublished Level 1 data used to calculate the EVL band irradiances \citep{woods_extreme_2012}, and thus give an answer that is up to 20\% lower for the 94 and 171 \AA\ channels. For this reason, we use the EVL data for all comparisons.

\subsection{Comparison with SORCE/SOLSTICE}
	\label{SS_aiasolstice}

\begin{figure}    
 \centerline{\includegraphics[width=0.7\textwidth,angle=0,clip=]{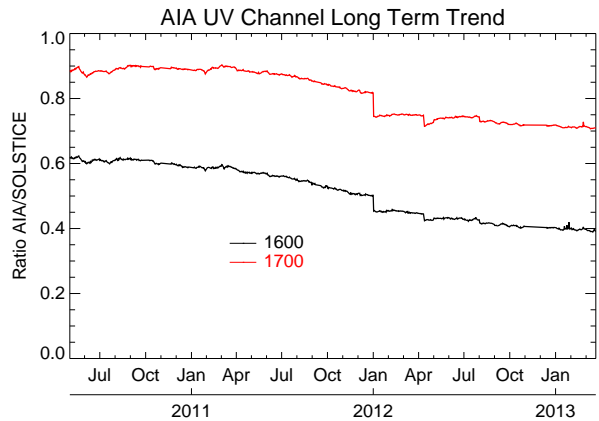}
              }
              \caption{The ratio of band irradiances measured in the AIA UV channels to those predicted using data from SORCE/SOLSTICE. As noted in \protect \citealp{boerner_initial_2012}, the absolute accuracy of the AIA UV channel calibration is not as good as for the EUV channels (a factor of 2 instead of 25\%). However, the trend plot shows only weak short- and long-term variation. (The step on 1 January 2012 is due to a change in the normalization of the AIA flatfield.)}
   \label{F_aia_solstice_trend}
\end{figure}

EVE does not cover the wavelength range of the AIA UV channels (1500--1800 \AA); however, SORCE/SOLSTICE \citep{mcclintock_solar-stellar_2005} measurements are available in this range. The approach described above can be used to fold SORCE/SOLSTICE data through the AIA UV channel response functions and compare the predicted and observed band irradiances for the 1600 and 1700 \AA\ channels. While the spectral resolution of SOLSTICE is roughly an order of magnitude worse than that of EVE, the AIA UV passbands are comparably broader than the EUV bands, and the solar spectrum in this range is less dominated by sharp lines, so the blurring of the spectrum by the instrumental response of SOLSTICE has a negligible impact on the predicted count rates. The results are shown in Figure \ref{F_aia_solstice_trend}. Again, low-order polynomials produce excellent fits to the observed trends with residuals $<$ 4\%. These fits are available through SSW.

\subsection{Comparison with Hinode/EIS}
	\label{SS_aiaeis}

The EIS instrument on {\it Hinode} \citep{culhane_euv_2007} is a slit spectrograph that operates in two EUV wavelength bands; the shorter band (170--210 \AA) completely overlaps the AIA 193 \AA\ channel. EIS offers excellent spectral resolution (approximately 50 m\AA), with a spatial resolution of 2 arcsec; it can be rastered to produce images with a field of view of 6 $\times$ 8.5 arcmin. While cross-calibration between EIS and EVE is difficult due to their discrepant fields of view, EIS has been cross-calibrated by the EUNIS sounding rocket \citep{wang_underflight_2011} and with SOHO/SUMER \citep{landi_relative_2010}.

In order to compare AIA and EIS observations, it is necessary to ensure that they are observing the same field. The EIS spectral data cube from a slit raster $I(\mathbf{x}, \lambda)$ is multiplied by the AIA 193 \AA\ response function $R(\lambda)$ and integrated over wavelength to produce a set of predicted 193 \AA\ pixel intensities $p_{\mathrm{pred}}(\mathbf{x})$. Then AIA 193 \AA\ images are used to build a ``simulated raster'' $p_{\mathrm{obs}}(\mathbf{x})$ such that each pixel in the result is chosen from an image taken at the same time as the corresponding EIS slit integration. The AIA/EIS normalization factor is then the ratio of $p_{\mathrm{obs}}/p_{\mathrm{pred}}$ for all points in the image. 

\begin{figure}    
 \centerline{\includegraphics[width=0.7\textwidth,angle=270,clip=]{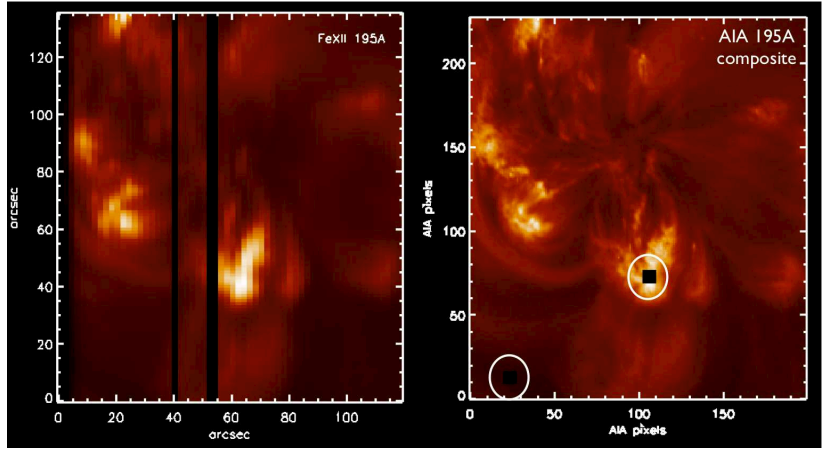}
              }
              \caption{(Left) A simulated AIA 193 \AA\ raster constructed by multiplying a 3D EIS spectral data cube with the AIA 193 \AA\ response function and integrating over wavelength. (Right) AIA 193 \AA\ observations of the same region; this is a pseudo-raster, as each pixel is chosen from an image taken at the same time as the corresponding EIS integration. The circled areas indicate the location of the bright ``moss'' and dim ``quiet Sun'' subregions selected for detailed comparison.}
   \label{F_eisraster}
\end{figure}

This technique was applied to an EIS raster taken in October 2010 (see Figure \ref{F_eisraster}). The field of view contained a small active region, including some moss, and some patches of quiet Sun. While the pixel-to-pixel variations in the AIA/EIS normalization factor could be substantial due to the difficulty in exactly co-aligning each pixel in space and time, the average over regions as small as $20 \times 20$ arcsec showed good agreement to within 15\% for the moss, quiet Sun, and the full field of view (see Table \ref{T_eiscount}). 

\begin{table}    
\caption{ EIS vs AIA 193 \AA\ channel.
}
\label{T_eiscount}
\begin{tabular}{rc}    
  \hline                 
Feature 		& AIA observed/EIS predicted	\\
  \hline
Moss			& 1.03	\\
Quiet Sun		& 0.98	\\
Full FOV		& 1.15	\\
  \hline
\end{tabular}
\end{table}

\section{Temperature Response}
	\label{S_tresp}

The temperature response function, $K(T)$, of an EUV narrowband imager is calculated from the wavelength response function and the plasma emissivity $G$:
\begin{equation} \label{E-tresp}
K_{i}(T) = \int_{0}^{\infty} G(\lambda, T) R_{i}(\lambda) \mathrm{d} \lambda .
\end{equation}
The emissivity is a description of the plama and atomic physics governing how material at a given temperature emits radiation. It includes empirically-derived values of the abundance of the various elements in the solar atmosphere, the ionization equilibrium of the ionic species of each element as a function of temperature, and the oscillator strengths of all the known emission lines of each ion (as well as a model of the continuum emission). This information is contained in the CHIANTI database, which represents a compendium of measurements and theoretical calculations of plasma properties.

Compiling the emissivity database and code is a challenging, ongoing research program, so the uncertainties associated with the emissivity are not negligible. For many of the emission lines targeted by AIA, the CHIANTI database is quite accurate; in particular, at wavelengths above the Al-L edge at 171 \AA\ , there have been numerous measurements of solar and stellar intensity, which have been used to refine the emissivity models (the same is true, to some extent, for the soft X-ray region between 6 and 50 \AA). However, prior to the launch of SDO, there had been very few measurements in the 50--150 \AA\ range, and as a result the emissivity in this range was not well characterized. 

\subsection{Benchmarking CHIANTI}
	\label{SS_chianti}

\begin{figure}    
 \centerline{\includegraphics[width=0.7\textwidth,angle=0,clip=]{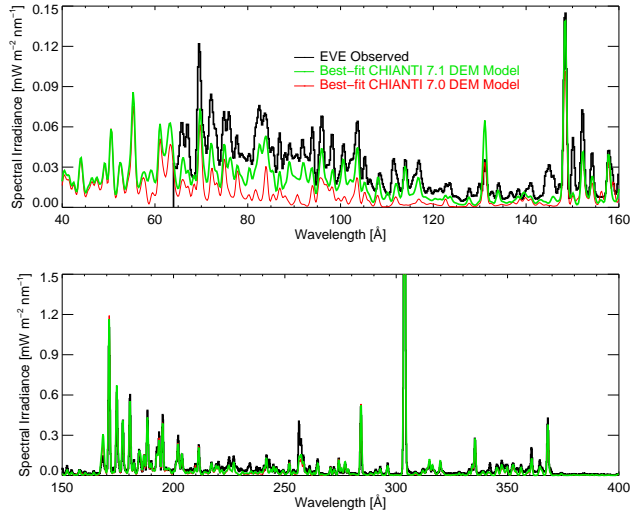}
              }
              \caption{An observed irradiance spectrum from SDO/EVE, compared with the best-fit spectrum using a DEM model and the CHIANTI database. The model accurately matches the observations in the 170--350 \AA\ range, but significantly underestimates the emission from 60--170 \AA\ .}
   \label{F_evechianti}
\end{figure}

Based on the observations of the 50--150 \AA\ range with EVE and the 94 and 131 \AA\ channels on AIA, it is clear that there are significant deficiencies in the spectral models in this wavelength range. Figure \ref{F_evechianti} shows an observed irradiance spectrum of the non-flaring Sun from EVE (black), along with a best-fit model spectrum generated using CHIANTI Version 7.0 (red) and 7.1 (green). The model shows excellent agreement with the many strong lines between 170--350 \AA\ (with the well-known exception of the 304 \AA\ He {\sc ii} line), implying that the assumptions about the thermodynamic state of the plasma is good. But between 50 and $\approx$ 150 \AA\, the model fails to reproduce the majority of the emission lines, and underpredicts the observed intensity by factors of 2--6. CHIANTI 7.1 clearly represents a substantial improvement, but there is still a significant amount of emission that is not accounted for. The missing flux is most significant in the quiet Sun; during flares, the emission in this wavelength range is dominated by a handful of strong lines (such as Fe {\sc xviii} 94 \AA\ and Fe {\sc xxi} 128 \AA\ imaged by AIA) that are well-reproduced by CHIANTI. However, the underestimate of the intensity from quiet Sun plasma can lead to false conclusions about the presence of hot plasma. 

This effect has been independently discovered by a number of authors ({\it e.g.} \citealp{aschwanden_solar_2011}, \citealp{teriaca_spectroscopic_2012}). In order to prove that this discrepancy is not a result of a calibration error in EVE, \citealp{testa_testing_2012} also examined spectra of Procyon taken by Chandra’s LETG. Again, they found that the CHIANTI model (which agrees well with the observed line intensities at more well-studied wavelengths) simply does not contain any information for many of the lines in this spectral range.

\subsection{Empirical Correction to AIA Temperature Response}
	\label{SS_chiantifix}

Work is in progress to update CHIANTI to include these missing lines (see, {\it e.g.}, \citealp{del_zanna_atomic_2012}); the release of Version 7.1 represents a major step. However, in the mean time, it is possible to make an empirical correction to the AIA temperature response functions themselves to attempt to account for the missing emission. This is done using the dataset of coordinated observations with AIA and EVE during a 1-h window around the X2 flare of 15 February 2011, and in samples of the irradiance taken daily throughout the SDO mission. (The flare spectra have a pre-flare baseline subtracted in order to isolate the dynamic hot component of the emission.) The EVE data are used to constrain a model of the DEM, as follows. The quiet Sun DEM derived by \citealp{dere_chianti_1997} from the observations of \citealp{vernazza_extreme_1978} is used as an initial guess, and parameterized as a cubic spline in $\log_{10}(T)$ using 4--6 spline knots. The DEM is combined with the emissivity function derived from CHIANTI to generate a synthetic spectrum, 
\begin{equation} \label{E-synth}
I(\lambda) = \int_{T_{\mathrm{min}}}^{T_{\mathrm{max}}} \! G(\lambda, T) \mathrm{DEM} (T)  \, \mathrm{d} T.
\end{equation}
The synthetic spectrum is blurred and resampled to EVE resolution as described in Section \ref{SS_aiaeve}, and the result is compared with the observed EVE Level 2 Version 2 spectrum in a set of windows 2 \AA\ wide centered on a set of strong emission lines in the spectral range where the CHIANTI model is known to be reasonably complete, and a $\chi^2$ is calculated by summing the squared differences of all EVE spectral bins in the selected windows. (Using windows rather than attempting to extract line intensities from the EVE measurements gives results that are more robust to blending that might result from EVE's moderate spectral resolution.) Note that the spectral windows around certain lines associated with high-temperature emission found in flares, including the Fe {\sc xviii} 94 \AA\ line and the Fe {\sc xxi} 128 \AA\ line imaged by AIA, are treated as upper limits and only factor into the $\chi^{2}$ when the predicted intensity exceeded the observed intensity; this allowed us to use those lines to constrain the hot end of the DEM during flares (since the CHIANTI data are fairly accurate for these hot lines), without being fooled by the deficiencies in the CHIANTI model of the adjacent cooler lines.

\begin{figure}    
 \centerline{\includegraphics[width=1.0\textwidth,angle=0,clip=]{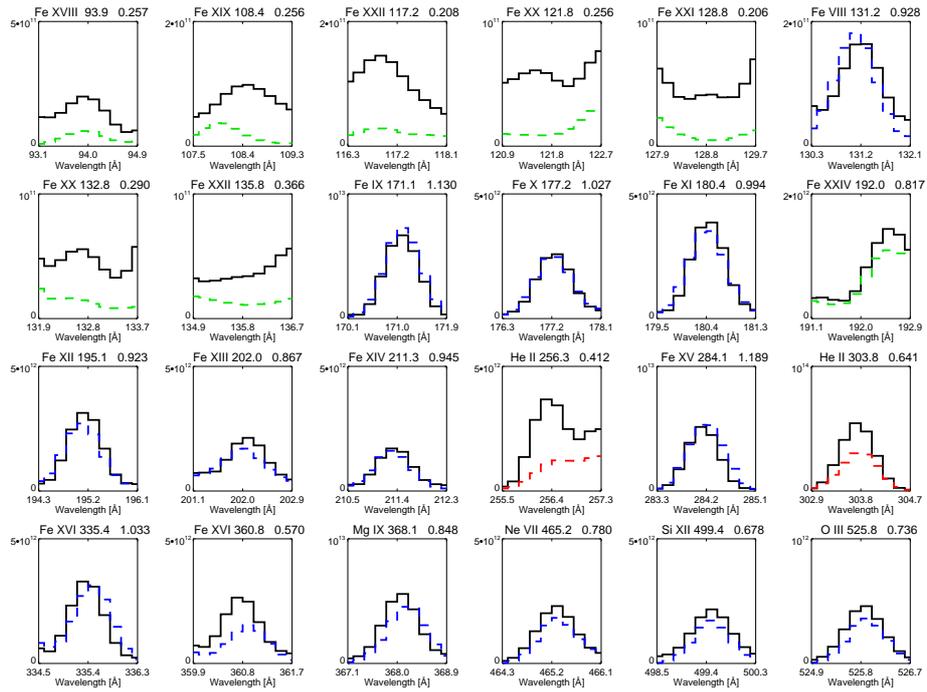}
              }
              \caption{A set of spectral windows around strong emission lines are used to constrain a DEM model of the EVE irradiance spectrum. The measured data are in black; colored curves indicate the model (blue shows data used in the fit; red curves are omitted from the fit, as the DEM model does not accurately represent the formation of these lines; green curves are used for lines treated as upper limits). For reference, the ratio of the intensity integrated over all bins in the window is listed in the title for each window (though this value is not used in the fit).}
   \label{F_evelines}
\end{figure}

The DEM spline knots were then adjusted iteratively using the Levenberg-Marquardt algorithm (the \verb+mpfit+ routine in IDL) to minimize the $\chi^{2}$. The DEM functions derived using this approach generally fit the EVE observations in the selected windows to better than 25\% (see Figure \ref{F_evelines}), so they can be considered a reasonably good representation of the thermal state of the corona. The comparison between the observed and best-fit synthetic spectra over the full EVE spectral range for both the daily sampled spectra and the X2 flare spectra (with the pre-flare spectrum subtracted) can be seen in movies posted at \url+http://www.lmsal.com/~boerner/crosscal/+. A number of characteristics of these movies are worth noting:

\begin{enumerate}

\item For the flare spectrum, the strong lines in the 94 and 131 \AA\ bands are fit quite well (the cooling of the flare from Fe {\sc xxi} to Fe {\sc xviii} is apparent). The 193 and 335 \AA\ bands also do reasonably well.

\item However, the 171 and 211 \AA\ channels do not match the preflare-subtracted observations. This is likely due to the fact that these channels do not have a significant contribution from hot lines, so the enhancement to their irradiance during the flare is negligible compared to fluctuations (or even dimmings; see, {\it e.g.}, \citealp{woods_new_2011}) in the global 1--2 MK corona; therefore, subtracting a static pre-flare background leaves only noise in these bands.

\item The daily samples (which typically resemble an average quiet-Sun DEM) generally show very good agreement in the range from 170--200 \AA\, including the lines not used in the fit.

\item There are some spectral ranges that are not well fit for the daily samples, including 200--250 \AA\ and 320--360 \AA. This is probably because the DEM is not well constrained below $\log_{10}(T)=5.6$ or so; however, this temperature range is not of primary significance for AIA.

\item Of course, the quiet Sun DEMs consistently underestimate the observations in the region from 60--150 \AA\, as expected (see Section \ref{SS_chianti}).

\end{enumerate}

Once we have determined DEM functions that accurately characterize the corona, we can adjust the AIA temperature response functions so that the count rate predicted by folding these DEMs through the response functions using Equation (\ref{E-dem}) matches the observed AIA band irradiance. 

The adjustment of the temperature response functions is a two-step process. Because we believe that CHIANTI accurately predicts the intensity of the hot lines that dominate during flares, any discrepancy between observed and DEM-predicted count rate using background-subtracted flare observations can be attributed to a normalization error in the temperature response function, and we can simply determine a scale factor $a_{0}$ that optimizes the agreement:

\begin{equation} \label{E-hotfit}
K_{\mathrm{scale}}(T) = a_{0} K_{\mathrm{orig}}(T) .
\end{equation}

\begin{figure}    
 \centerline{\includegraphics[width=0.7\textwidth,angle=0,clip=]{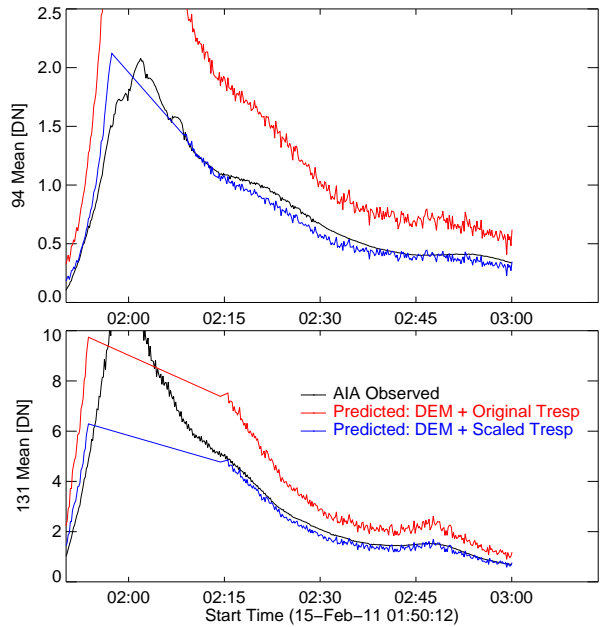}
              }
              \caption{The DEMs derived from the flare observations are used to normalize the temperature response functions. The irradiance predicted using the DEMs and the temperature response functions with the best-fit normalization constant closely matches the observed irradiance during the cooling phase of the flare.}
   \label{F_flarenorm}
\end{figure}

The results for the 94 and 131 \AA\ channels are shown in Figure \ref{F_flarenorm}. Note that the 12-min period around the peak of the flare is omitted from the fit because substantial saturation in the AIA image reduces the reliability of the AIA irradiance measurements. The band irradiance predicted using the scaled temperature response functions match the observations very closely. The best-fit scale factors are 0.62 for the 94 channel ( {\it i.e.} the count rates are only 62\% of what would be predicted using the nominal temperature response function and the best-fit flare DEMs), and 0.63 for the 131 \AA\ channel.

We then compare this scale factor derived in temperature space with the $F_{\mathrm{norm}}$ derived in ``wavelength space'' using the same dataset (by folding the preflare-subtracted EVE spectral irradiance through the wavelength response function, as in Section \ref{SS_aiaeve}). For the 131 \AA\ channel, the wavelength comparison suggests a correction factor of 0.64, which is quite close to what we find in temperature space. However, we note that the wavelength space comparison gives a correction factor of 0.81 when we look at the spectrum before and after the flare, without subtracting the baseline. We interpret this to mean that the effective area of the 131 \AA\ channel needs to be scaled by 0.64 at the wavelength of the Fe {\sc xxi} flare line, but only by 0.81 at the wavelength of the Fe {\sc viii} line that dominates in non-flaring conditions. Rather than attempt to adjust the shape of the wavelength response function, we adjust the entire response function by 0.81 to agree with the wavelength cross-calibration during non-flaring times, and then apply an additional scale factor of 0.79 to the portion of the temperature response function above 6.7 in $\log_{10}(T)$.
 
For the 94 channel, the correction derived in wavelength space is close to 1.0, rather than the 0.62 derived in temperature space. Most of the discrepancy can be attributed to the wavelength resolution effect noted above; if we take the synthetic spectrum predicted by the best-fit flare DEMs and blur it to EVE's spectral resolution, the predicted count rates in the 94 \AA\ channel are approximately 30\% lower than the predictions obtained with the unblurred spectrum. The remaining 8\% discrepancy may be attributable to errors in the DEM fit. However, note that the adjustment to the high-temperature component of the 94 channel temperature response derived from this comparison is likely to be more accurate than the adjustment implied by, and could not be obtained directly from, folding the EVE observations through the wavelength response function. Therefore, we scale the entire 94 \AA\ channel response down by 0.7.

Having fixed the normalization of the responses such that they give excellent agreement with EVE spectra and with EVE-constrained DEMs during flares, the next step is to add come contribution to the lower-temperature portion of the functions so that the daily sample DEMs accurately predict the observations.
\begin{equation} \label{E-coolfit}
K_{\mathrm{fit}}(T) = K_{\mathrm{scale}}(T) + \sum_{n=1}^2 a_{n} G_{n}(T).
\end{equation}
The shape of the contribution functions $G_{n}(T)$ are chosen based on estimates of the temperature characteristics of the emission missing from each bandpass, derived either from surveys of the atomic databases (see, {\it e.g.}, \citealp{del_zanna_benchmarking_2012} and following, who note that there are likely to be strong Fe {\sc ix} lines missing from the 94 \AA\ channel), or from comparing the morphology of structures seen in the images to images from lines at well-known temperature (\citealp{warren_observations_2011-1} note that, in the quiet Sun, the 94 \AA\ images most closely resemble EIS and AIA Fe {\sc xii} images). The $a_{n}$ coefficients are then found by minimizing the $\chi^{2}$. For the 94 \AA\ channel, we chose $G_{1}(T)$ to be the temperature distribution of the Fe {\sc ix} line at 171 \AA\, and $G_{2}(T)$ to be the shape of the Fe {\sc xii} 195 \AA\ line. For the 131 \AA\ channel, $G_{1}(T)$ was based on the 180 \AA\ Fe {\sc xi} line, and $G_{2}(T)$ was the shape of the Fe {\sc viii} line already in the 131 \AA\ band. Alternate parameterizations were tried, with $n=1$ to $n=3$ and different temperature lines added to each band. The results are not very sensitive to the exact details of the added contribution; for example, agreement between predicted and observed counts in the 131 \AA\ channel would not be very different if we chose to add an Fe {\sc x}-like component instead of an Fe {\sc xi}-like component, and the relative balance of Fe {\sc ix} and Fe {\sc xii} added to the 94 \AA\ channel is not extremely well constrained. However, the basic shape of the corrections is well-motivated and provides very good agreement with observations.

\begin{figure}    
 \centerline{\includegraphics[width=0.9\textwidth,angle=0,clip=]{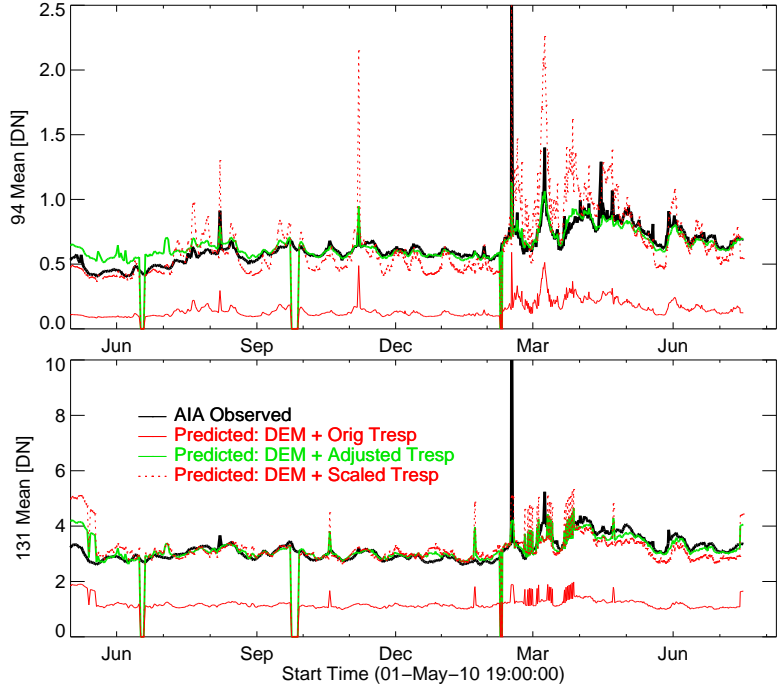}
              }
              \caption{Daily samples of the EUV irradiance taken over a broad range of solar conditions were used to constrain the shape of the cool end of the temperature response functions. The original temperature response functions (red) underestimate the observations (black) by factors of 2--4. Contributions from Fe {\sc viii}--{\sc xii} were added until the agreement between the observed band irradiance (black) and the count rate predicted using the best-fit DEM and the modified temperature response (green) matched the magnitude and the variation of the observations. Simply scaling up the cool portion of the original temperature response function by a best-fit factor (red dashed line) matches the average value of the signal, but not the details of its variation.}
   \label{F_coolfit}
\end{figure}

\begin{figure}    
 \centerline{\includegraphics[width=0.8\textwidth,angle=0,clip=]{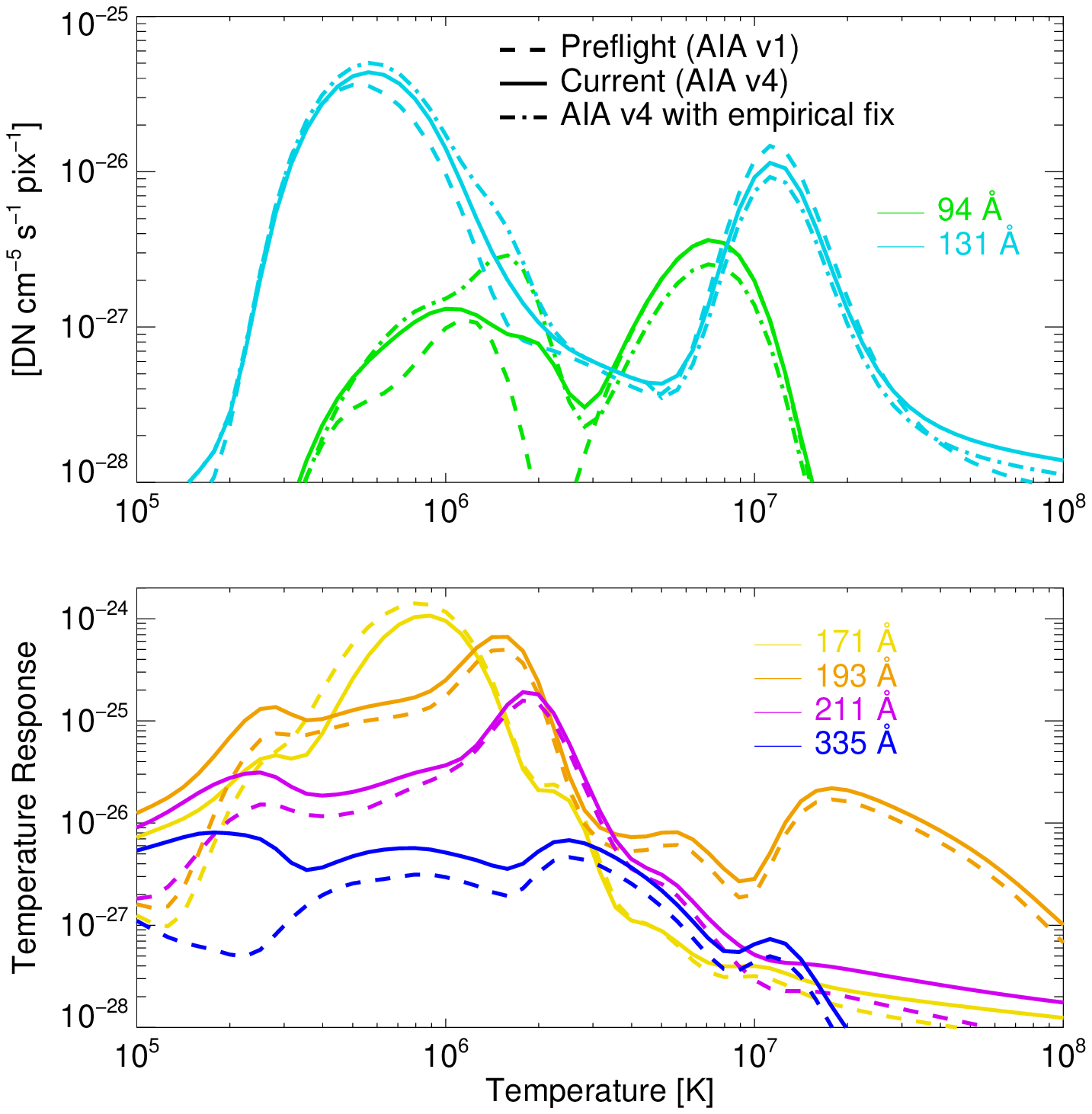}
              }
              \caption{The temperature response functions for the AIA EUV channels with the corrections discussed here applied. The pre-flight calculation (using ground calibration of the effective area combined with atomic data from CHIANTI Version 6.0.1) is shown with a dashed line (this is Version 1 of the AIA calibration). The updated temperature response calculated by cross-calibration of the wavelength response function with EVE combined with atomic data from CHIANTI Version 7.1 is shown with the solid lines. In the top panel, the empirical correction to the 131 and 94 \AA\ channels is also shown with a dash-dotted line. For both channels, the high-temperature peak is slightly reduced, and there is significant additional contribution from material around 1 MK.}
   \label{F_fit_tresps}
\end{figure}

That agreement is shown in Figure \ref{F_coolfit}. The observed band irradiances are plotted in black, and the predictions given the best-fit DEMs and the original temperature response functions $K_{\mathrm{orig}}(T)$ are shown in red, while the predictions obtained with the best-fit response functions $K_{\mathrm{fit}}(T)$ are in green. Note that the predictions obtained by simply scaling up the cool end of the temperature response function (as was done in \citealp{aschwanden_solar_2011}), plotted with dotted red lines, improve the agreement substantially, but clearly do not match the detailed behavior of the observations as well as the best-fit modifications, especially in the 94 \AA\ channel. The best-fit response functions are shown in Figure \ref{F_fit_tresps}. 

As noted in Section \ref{SS_chiantifix}, Version 7.1 of CHIANTI (released in October 2012) added a large number of emission lines in the 50--160 \AA\ range and thus reduced the need for and the impact of the empirical correction to the AIA temperature response. The AIA response functions were updated to Version 4 to incorporate these new emission lines, with the empirical correction (accessible with the \verb+chiantifix+ keyword to the \verb+aia_get_response+ function) retuned appropriately. The history of the AIA calibration versions is summarized in Table \ref{T_calversion}.

\begin{table}    
\caption{ AIA calibration version history
}
\label{T_calversion}
\begin{tabular}{ccccccc}    
  \hline                 
AIA Calibration	& Release	& CHIANTI	& \multicolumn{4}{ c }{Approx. scale of empirical fix} \\
version 		& date		& version	& Hot 94 & Cool 94 & Hot 131 & Cool 131 \\
  \hline
1 				& Aug 2010	& 6.0.1		& -- & -- & -- & -- \\
2				& Feb 2012	& 7.0		& 0.55 & 4.0 & 0.85 & 2.0 \\
3				& Sep 2012	& 7.0		& 0.55 & 4.0 & 0.85 & 2.0 \\
4				& Feb 2013	& 7.1		& 0.70 & 2.0 & 0.79 & 1.0 \\
  \hline
\end{tabular}
\end{table}

\section{Implications for Thermal Analysis}
	\label{S_dem}

In order to validate these results on a separate set of observations, and to characterize their effect on the conclusions obtained from thermal analysis with AIA, we carried out a series of inversions using both the original and the modified temperature response functions. 

\subsection{DEMs with AIA Alone}
	\label{SS_aia_dem}
	
For the first of these, we used only AIA data. The six Fe channels of AIA can provide reliable temperature constraints with moderate resolution (0.3 in $\log_{10}(T)$) for optically thin plasma in the range of 0.7--3 MK \citep{guennou_accuracy_2012}. Averaging over large regions of the corona above the limb during non-flaring conditions therefore provides an effective benchmark for DEM inversions. We divided the off-limb corona from the period prior to the X2.2 flare on 15 February 2011 into 25 equally-sized sectors, and integrated the signal in the six Fe channels from each sector. 

\begin{figure}    
 \centerline{
 			\includegraphics[width=0.43\textwidth,angle=0,clip=]{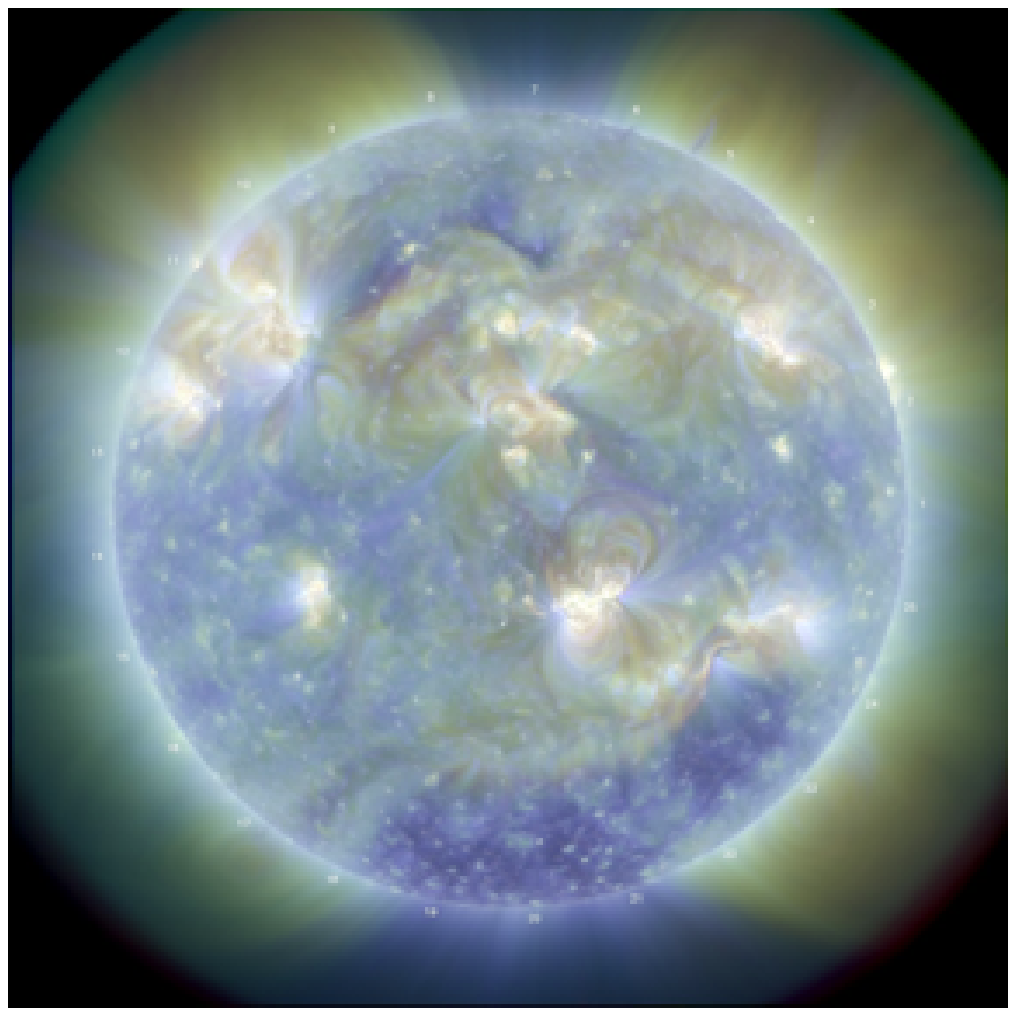}
    		\includegraphics[width=0.57\textwidth,angle=0,clip=]{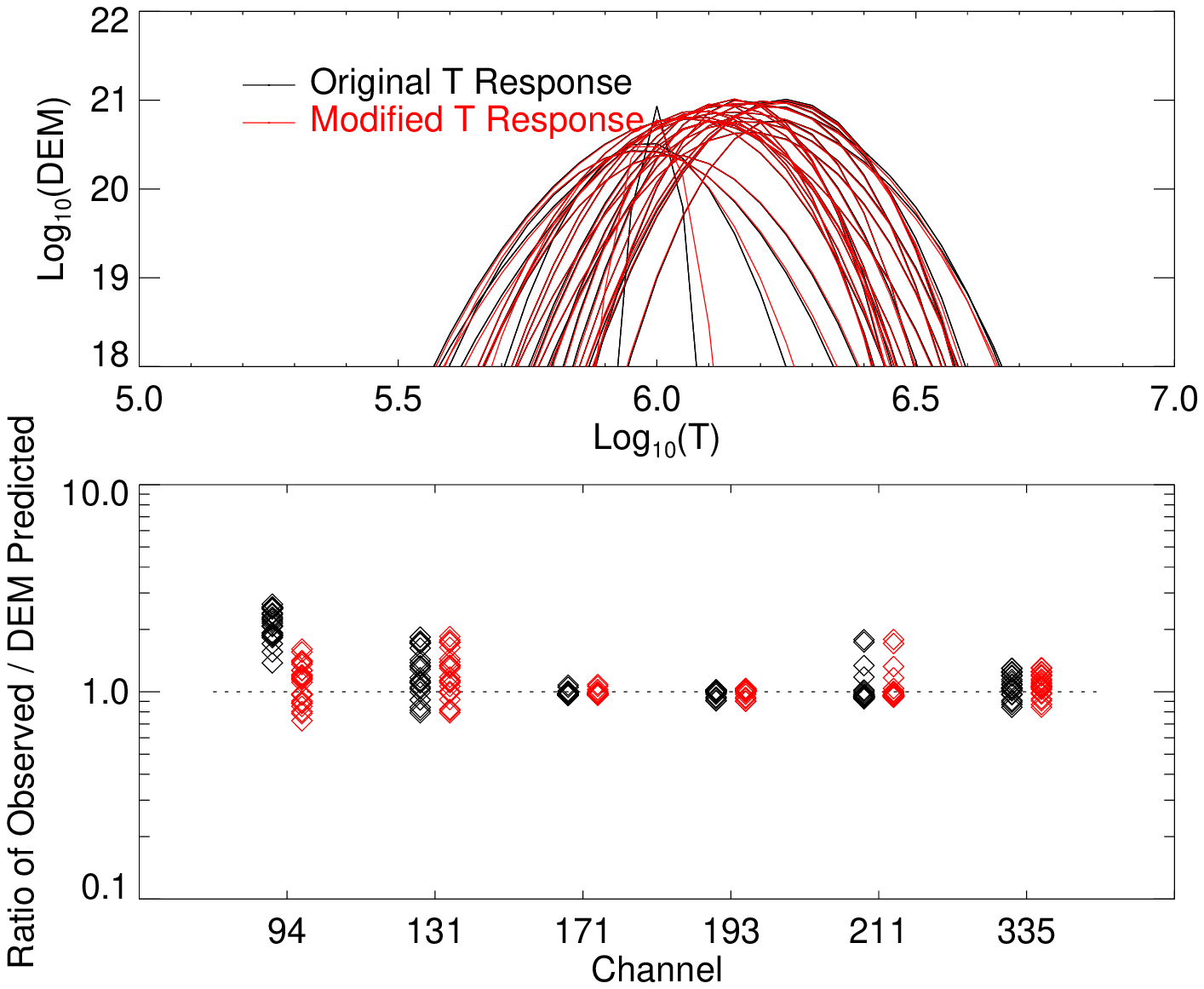}
              }
              \caption{The region above the solar limb was divided into 25 sectors (a). The best-fit Gaussian DEMs are shown in (b), along with the ratio of the observed count rate in each channel to that predicted by the DEM (c). The modified temperature response functions produce much better agreement in the 94 and 131 channels.}
   \label{F_offlimb}
\end{figure}

For each sector, a DEM inversion was performed using a single Gaussian function of temperature, with both the original (black in Figure \ref{F_offlimb}) and modified (red in Figure \ref{F_offlimb}) temperature response functions. Because the 171, 193 and 211 \AA\ channels are an order of magnitude more sensitive to plasma at the temperature of the quiescent corona, their signals dominate the fit. The recovered DEM functions show only minor differences when the modified 94 and 131 \AA\ responses are used, generally producing slightly narrower gaussians. However, the modified responses dramatically improve the agreement with the 94 and 131 \AA\ observations. With the original response functions, the gaussian DEMs underpredict the flux in both channels by the same factor of 2--4 noted with DEMs derived from EVE. This result further validates the corrections we derived from comparison with EVE.
 
Without the inclusion of the cooler contributions in the 94 and 131 \AA\ response functions, the only way to explain the observed signal in the 94 and 131 channels would be to assume that a substantial amount of hot ($T>6$ MK) plasma exists throughout the corona. The most significant impact of the modification to the temperature response functions is the suppression of spurious hot tails on the inferred DEM distributions.

\subsection{DEMs with AIA and EIS + XRT}
	\label{SS_aia_eis_dem}

A secondary benefit of ensuring accurate photometric calibration is that it allows us to leverage observations from multiple instruments. Combining AIA data with observations from EIS, as in \citealp{warren_constraints_2011}, makes it possible to measure temperatures with finer coverage and resolution than with AIA alone, and to take advantage of the diagnostic line ratios in the EIS data set. Adding in data from {\it Hinode}/XRT allows further insight, in particular by constraining the high-temperature end of the temperature distribution \citep{winebarger_using_2011}.

\begin{figure}    
 \centerline{\includegraphics[width=0.7\textwidth,angle=0,clip=]{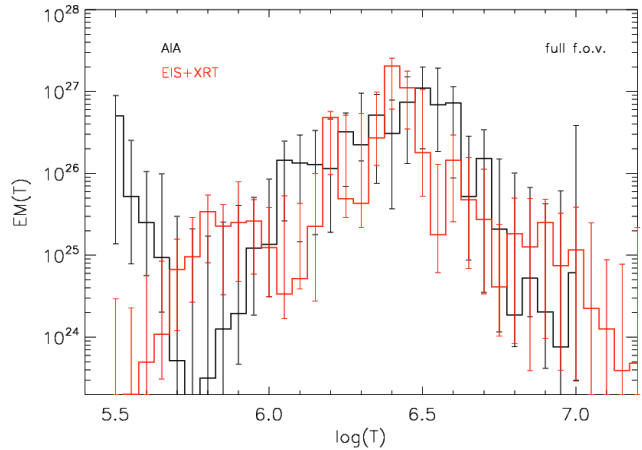}
              }
              \caption{{\it Hinode}/EIS and XRT were used to derive DEMs for the sub-regions shown in Figure \ref{F_eisraster}; those DEMs are fairly similar to those obtained with the six AIA Fe channels alone in the temperature range where most of the AIA emission is formed.}
   \label{F_eis_aia_dem}
\end{figure}

Using the observations from Figure \ref{F_eisraster}, we fit DEMs for the subregions identified in Table \ref{T_eiscount} using data from AIA alone, and with a combination of EIS and XRT. The results are shown in Figure \ref{F_eis_aia_dem}. As should be expected, the combination of the large number of EUV lines from EIS and the high-temperature constraint from XRT provides the most complete temperature constraint; however, the agreement between the AIA-only DEM and the one obtained from EIS and XRT is reasonably good, especially within the temperature range from 1--4 MK where the AIA channels are most sensitive.

\begin{figure}    
 \centerline{\includegraphics[width=0.7\textwidth,angle=0,clip=]{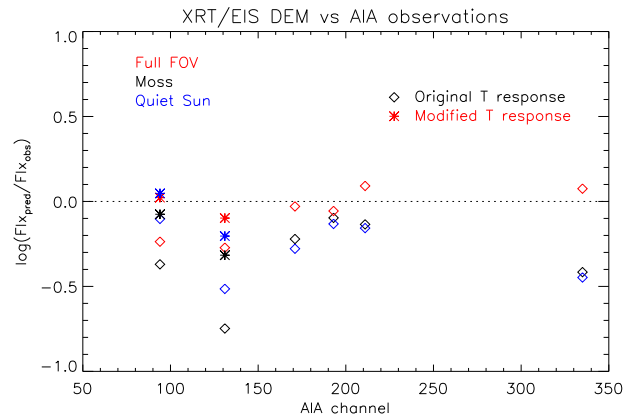}
              }
              \caption{The DEMs obtained with EIS and XRT (see Figure \ref{F_eis_aia_dem}) were then folded through the AIA temperature response functions to predict count rates for those regions. The observed AIA count rates agree better with those predicted using the modified response functions for the 94 and 131 \AA\ channels.}
   \label{F_eis_dem_counts}
\end{figure}

In order to further validate the modifications to the 94 and 131 \AA\ response functions, we then used the DEM inferred from EIS and XRT observations to predict AIA count rates using both the original and the modified temperature response functions. The results are shown in Figure \ref{F_eis_dem_counts}. Once again, the agreement in the 94 and 131 \AA\ channels is dramatically improved with the revised functions. Also, the fact that the EIS/XRT-derived DEM agrees as well as it does with the AIA observations emphasizes the fact that the apparent fine-scale discrepancies between the DEMs shown in Figure \ref{F_eis_aia_dem} are not necessarily significant. AIA data alone would not reject a DEM solution like the one produced with EIS and XRT.

\section{Conclusions}
	\label{S_conclusions}

The photometric calibration of SDO/AIA as a function of wavelength shows generally good agreement with SDO/EVE, {\it Hinode}/EIS, and SORCE/SOLSTICE. If we assume that the calibration of EVE is correct, we can correct for residual errors in the AIA calibration and ongoing changes in the instrument sensitivity by normalizing the AIA wavelength response functions using EVE observations. However, there is still some uncertainty in the shape of the 335 \AA\ passband, which cannot be corrected with a simple normalization.

The determination of the instrument response as a function of temperature is limited by the deficiency of the CHIANTI database in the 50--170 \AA\ wavelength range; however, pending improvements to CHIANTI, we propose an empirical correction to the temperature response functions of the 94 and 131 \AA\ channels that produces good agreement with DEM models obtained from other sources.

These improvements to the accuracy of the AIA response functions allow more accurate quantitative analysis of the data obtained by AIA.

\begin{acks}
The authors thank the members of the EVE team for providing helpful advice and excellent data. This work is supported by NASA under contract NNG04EA00C.
\end{acks}

\mbox{}~\\

\end{article} 

\end{document}